\documentclass[acmtog]{acmart}

\usepackage{booktabs} 

\setcitestyle{square}

\usepackage{soul}
\usepackage{multirow}
\usepackage[ruled]{algorithm2e} 

\SetAlFnt{\small}
\SetAlCapFnt{\small}
\SetAlCapNameFnt{\small}
\SetAlCapHSkip{0pt}
\IncMargin{-\parindent}

\received{January 2019}

\definecolor{red}{rgb}{0.8,0,0}
\definecolor{purered}{rgb}{1,0,0}
\definecolor{darkred}{rgb}{0.6,0,0}
\definecolor{green}{rgb}{0.0,0.5,0}
\definecolor{blue}{rgb}{0,0,0.75}
\definecolor{darkblue}{rgb}{0,0,0.55}
\definecolor{orange}{rgb}{0.9,0.3,0.1}
\definecolor{purple}{rgb}{0.6,0.0,0.6}
\definecolor{cyan}{rgb}{0.0,0.7,0.7}
\definecolor{darkgray}{rgb}{0.4,0.4,0.4}
\definecolor{bronze}{rgb}{0.8, 0.5, 0.2}
\definecolor{dorange}{rgb}{0.75, 0.4, 0.0}

\begin{document}
	
\sloppy

\acmJournal{TOG}
\acmYear{2019}

\settopmatter{printacmref=false} 
\renewcommand\footnotetextcopyrightpermission[1]{} 
\pagestyle{plain}
\makeatletter
\renewcommand\@formatdoi[1]{\ignorespaces}
\makeatother

\title{A Similarity Measure for Material Appearance}

\author{Manuel Lagunas}
\affiliation{%
  \institution{Universidad de Zaragoza, I3A}
  \country{Spain}
  \city{Zaragoza}
}
\email{mlagunas@unizar.es} 

\author{Sandra Malpica}
\affiliation{%
  \institution{Universidad de Zaragoza, I3A}
  \country{Spain}
  \city{Zaragoza}
}
\email{smalpica@unizar.es} 

\author{Ana Serrano}
\affiliation{%
  \institution{Universidad de Zaragoza, I3A}
  \country{Spain}
  \city{Zaragoza}
}
\email{anase@unizar.es} 

\author{Elena Garces}
\affiliation{%
  \institution{Universidad Rey Juan Carlos}
  \country{Spain}
  \city{Madrid}
}
\email{elena.garces@urjc.es} 

\author{Diego Gutierrez}
\affiliation{%
  \institution{Universidad de Zaragoza, I3A}
  \country{Spain}
  \city{Zaragoza}
}
\email{diegog@unizar.es} 

\author{Belen Masia}
\affiliation{%
  \institution{Universidad de Zaragoza, I3A}
  \country{Spain}
  \city{Zaragoza}
}
\email{bmasia@unizar.es} 

\begin{teaserfigure}
	\centering
	\includegraphics[width=\columnwidth]{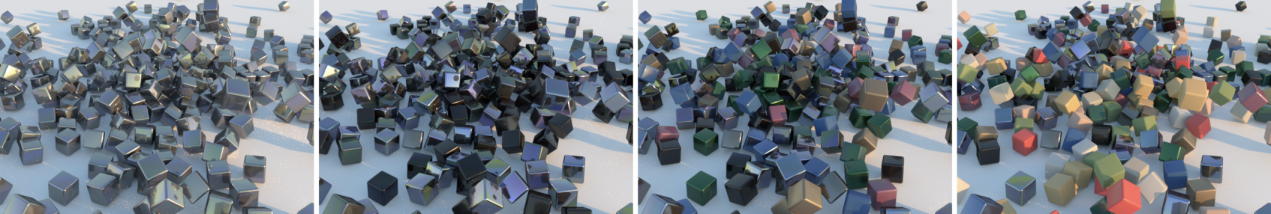}
	  	\caption{The cubes in the leftmost image have all been rendered with the same aluminium material. Our similarity measure for material appearance can be used to automatically generate alternative depictions of the same scene, where the similarity of the materials varies in a controlled manner. The next three images show results with materials randomly chosen by progressively extending the search distance from the original aluminium, from similar in appearance to farther away materials within the same dataset.} 
	\label{fig:teaser}
\end{teaserfigure}

\begin{abstract}
We present a model to measure the similarity in appearance between different materials, which correlates with human similarity judgments. We first create a database of 9,000 rendered images depicting objects with varying materials, shape and illumination. We then gather data on perceived similarity from crowdsourced experiments; our analysis of over 114,840 answers suggests that indeed a shared perception of appearance similarity exists. We feed this data to a deep learning architecture with a novel loss function, which learns a feature space for materials that correlates with such perceived appearance similarity. Our evaluation shows that our model outperforms existing metrics. Last, we demonstrate several applications enabled by our metric, including appearance-based search for material suggestions, database visualization, clustering and summarization, and gamut mapping. 
\end{abstract}


%
%


\maketitle

\newcommand{\mparagraph}[1]{\vspace{1mm}\noindent\textbf{#1.\hspace{2mm}}}

\section{Introduction}


Humans are able to recognize materials, compare their appearance, or even infer many of their key properties effortlessly, just by briefly looking at them. Many works propose classification techniques, although it seems clear that labels do not suffice to capture the richness of our subjective experience with real-world materials~\cite{fleming2017}. Unfortunately, the underlying perceptual process of material recognition is complex, involving many distinct variables; such process is not yet completely understood~\cite{anderson2011,fleming2014,maloney2010}.  


Given the large number of parameters involved in our perception of materials, many works have focused on individual attributes (such as the perception of gloss ~\cite{pellacini2000,wills2009}, or translucency~\cite{Gkioulekas}), while others have focused on particular applications like material synthesis~\cite{Zsolnai2018}, editing~\cite{serrano2016}, or filtering~\cite{jarabo2014btf}. However, the fundamentally difficult problem of establishing a \textit{similarity measure for material appearance} remains an open problem. 
Material appearance can be defined as ``the visual impression we have of a material''~\cite{dorsey2010digital}; as such, it depends not only on the BRDF of the material, but also on external factors like lighting or geometry, as well as human judgement~\cite{fleming2014,adelson2001seeing}. This is different from the common notion of image similarity (devoted to finding detectable differences between images, e.g.,~\cite{wang2004image}), or from similarity in BRDF space (which has been shown to correlate poorly with human perception, e.g.,~\cite{serrano2016}).
Given the ubiquitous nature of photorealistic computer-generated imagery, and emerging fields like computational materials, a similarity measure of material appearance could be valuable for many applications.

Capturing a human notion of perceptual similarity in different contexts has been an active area of research recently~\cite{GarcesSIG2014,agarwal2007generalized,lun2015elements}. In this paper we 
develop a novel image-based material appearance similarity measure derived from a learned feature space. 
This is challenging, given the subjective nature of the task, and the interplay of confounding factors like geometry or illumination in the final perception of appearance.
Very recent work suggests that perceptual similarity may be an emergent property, and that deep learning features can be trained to learn a representation of the world that correlates with perceptual judgements~\cite{zhang2018}. 
Inspired by this, we rely on a combination of large amounts of images, crowdsourced data, and deep learning. %
In particular, we create a diverse collection of 9,000 stimuli using the measured, real-world materials in the MERL dataset~\cite{matusik2003}, which covers a wide variety of isotropic appearances, and a combination of different shapes and environment maps. 
Using triplets of images, we gather information through Mechanical Turk, where participants are asked which of two given examples has a more similar appearance to a reference. Given our large stimuli space, we employ an adaptive sampling scheme to keep the number of triplets manageable. 
From this information, we learn a model of material appearance similarity using a combined loss function that enforces learning of the appearance similarity information collected from humans, and the main features that describe a material in an image; this allows us to learn the notion of material appearance similarity explained above, dependent on both the visual impression of the material, and the actual physical properties of it.

To evaluate our model, we first confirm that humans do provide reliable answers, suggesting a shared perception of material appearance similarity, and further motivating our similarity measure. We then compare the performance of our model against humans: Despite the difficulty of our goal, our model performs on par with human judgements, yielding results better aligned with human perception than current metrics. 
Last, we demonstrate several applications that directly benefit from our metric, such as material suggestions, database visualization, clustering and summarization, or gamut mapping.
In addition to the 9,000 rendered images, our database also includes surface normals, depth, transparency, and ambient occlusion maps for each one, while our collected data contains 114,840 answers; we provide both, along with our pre-trained deep learning framework, in order to help future studies on the perception of material appearance\footnote{http://webdiis.unizar.es/\textasciitilde
mlagunas/publication/material-similarity/}.

\section{Related Work}
	
\mparagraph{Material perception} 
There have been many works aiming to understand the perceptual properties of BRDFs~\cite{anderson2011,fleming2015, fleming2014, maloney2010}; a comprehensive review can be found in the work of Thompson and colleagues~\shortcite{Thompson2011}. Finding a direct mapping between perceptual estimates and the physical material parameters is a hard task involving many dimensions, not necessarily correlated. Many researchers focus on the perception of one particular property of a given material (such as glossiness \cite{chadwick2015, pellacini2000,wills2009}, translucency~\cite{Gkioulekas, gkioulekas2013understanding}, 
or viscosity~\cite{van2018}), or one particular application (such as filtering~\cite{jarabo2014btf}, computational aesthetics~\cite{Cunningham2007}, or editing~\cite{serrano2016,mylo2017appearance}). Leung and Malik~\shortcite{leung2001representing} study the appearance of flat surfaces based on textural information. Other recent works analyze the influence on material perception of external factors such as illumination~\cite{ho2006,vangorp2017perception,Krivanek2010}, motion~\cite{doerschner2011}, or shape~\cite{vangorp2007,havran2016}.

A large body of work has been devoted to analyzing the relationships between different materials, and deriving low-dimensional perceptual embeddings~\cite{matusik2003,wills2009,serrano2016,soler2018}. These embeddings can be used to derive material similarity metrics, which are useful to determine if two materials convey the same appearance, and thus benefit a large number of applications (such as BRDF compression, fitting, or gamut mapping). A number of works have proposed different metrics, computed either directly over measured BRDFs~\cite{fores2012,ngan2005}, in image space~\cite{pereira2012,ngan2006,sun2017}, or in reparametrizations of BRDF spaces based on perceptual traits~\cite{pellacini2000, serrano2016}. Our work is closer to the latter; however, rather than analyzing perceptual traits in isolation, we focus on the overall appearance of materials, and derive a similarity measure that correlates with the notion of material similarity as perceived by humans.

\mparagraph{Learning to recognize materials}    
Image patches have been shown to contain enough information for material recognition~\cite{schwartz2018}, and several works have leveraged this to derive learning techniques for material recognition tasks.     
Bell et al.~\shortcite{bell2015material} introduce a CNN-based approach for local material recognition using a large annotated database, while Schwartz and Nishino explicitly introduce global contextual cues~\shortcite{schwartz2016}. 	
Other works add more information such as known illumination, depth, or motion. Georgoulis et al.~\shortcite{georgoulis2017} use both an object's image and its geometry to create a full reflectance map, which is later used as an input to a four-class coarse classifier (metal, paint, plastic or fabric). For a comprehensive study on early material recognition systems and latest advances, we refer to the reader to the work of Fleming~\shortcite{fleming2017}. These previous works focus mainly on classification tasks, however \emph{mere labels do not capture the richness of our subjective experience of materials in the real world}~\cite{fleming2017}. 

Recent work has shown the extraordinary ability of deep features to match human perception in the assessment of perceptual similarity between two images~\cite{zhang2018}. Together with the success of the works mentioned above, this provides motivation for the combination of user data and deep learning that we propose in this work.

\mparagraph{Existing datasets} 
Early image-based material datasets include CURet~\cite{Dana1999}, KTH-TIPS~\cite{Hayman2004}, or FMD~\cite{sharan2009}. OpenSurfaces~\cite{bell2013opensurfaces} includes over 20,000 real-world images, with surface properties annotated via crowdsourcing. This dataset has served as a baseline to others, such as the Materials in Context Database (MINC) \cite{bell2015material}, an order of magnitude larger;  SynBRDF~\cite{kim2017lightweight}, with 5,000 rendered materials randomly sampled from OpenSurfaces; or MaxBRDF dataset~\cite{VCGL19}, which includes synthetic anisotropic materials.    

Databases with \emph{measured} materials include MERL~\cite{matusik2003} for isotropic materials, UTIA~\cite{filip2014template} for anisotropic ones, the  Objects under Natural Illumination Database~\cite{lombardi2012reflectance}, which includes calibrated HDR information, or the recent, on-going database by Dupuy and Jakob which measures spectral reflectance~\shortcite{dupuy2018}. 
    We choose as a starting point the MERL dataset, since it contains a wider variety of isotropic materials, and it is still being successfully used in many applications such as gamut mapping~\cite{sun2017}, material editing~\cite{serrano2016,Sun2018}, BRDF parameterization~\cite{soler2018}, or photometric light source estimation~\cite{lu2018}.

\section{Materials dataset}
\label{sec:dataset}

\subsection{Why a new materials dataset?}

To obtain a meaningful similarity measure of material appearance we require a large dataset with the following characteristics: 

\begin{itemize}
\item Data for a wide variety of materials, shapes, and illumination conditions.
\item Samples featuring the \emph{same} material rendered under different illuminations and with different shapes.
\item Materials represented by measured BRDFs, with reflectance data captured from real materials.
\item Real-world illumination, as given by captured environment maps.
\item A large number of samples, amenable to learning-based frameworks.
\end{itemize}

These characteristics enable renditions of complex, realistic appearances and will be leveraged to train our model, explained in Section~\ref{sec:cnn}. To our knowledge, none of the existing material datasets features all these characteristics.

\subsection{Description of the dataset}

In the following, we briefly describe the characteristics of our dataset, and refer the reader to the supplementary material for further details. 

\mparagraph{Materials} 
Our dataset includes all 100 materials from the MERL BRDF database~\cite{matusik2003}. This database was chosen as starting point since it includes real-world, measured reflectance functions covering a wide range of reflectance properties and types of materials, including paints, metals, fabrics, or organic materials, among others. 

\mparagraph{Illumination} 
We use six natural illumination environments, since they are favored by humans in material perception tasks~\cite{fleming2003real}. They include a variety of scenes, ranging from indoor scenarios to urban or natural landscapes, as high-quality HDR environment maps\footnote{ http://gl.ict.usc.edu/Data/HighResProbes/}.

\mparagraph{Scenes} 
Our database contains thirteen different 3D models, with an additional camera viewpoint for two of them, defining our fifteen scenes. 
It includes widely used 3D models, and objects that have been specifically designed for material perception studies~\cite{havran2016,vangorp2007}. The viewpoints have been chosen to cover a wide range of features such as varying complexity, convexity, curvature, and coverage of incoming and outgoing light directions.

\begin{figure}
	\centering
	\includegraphics[width=\columnwidth]{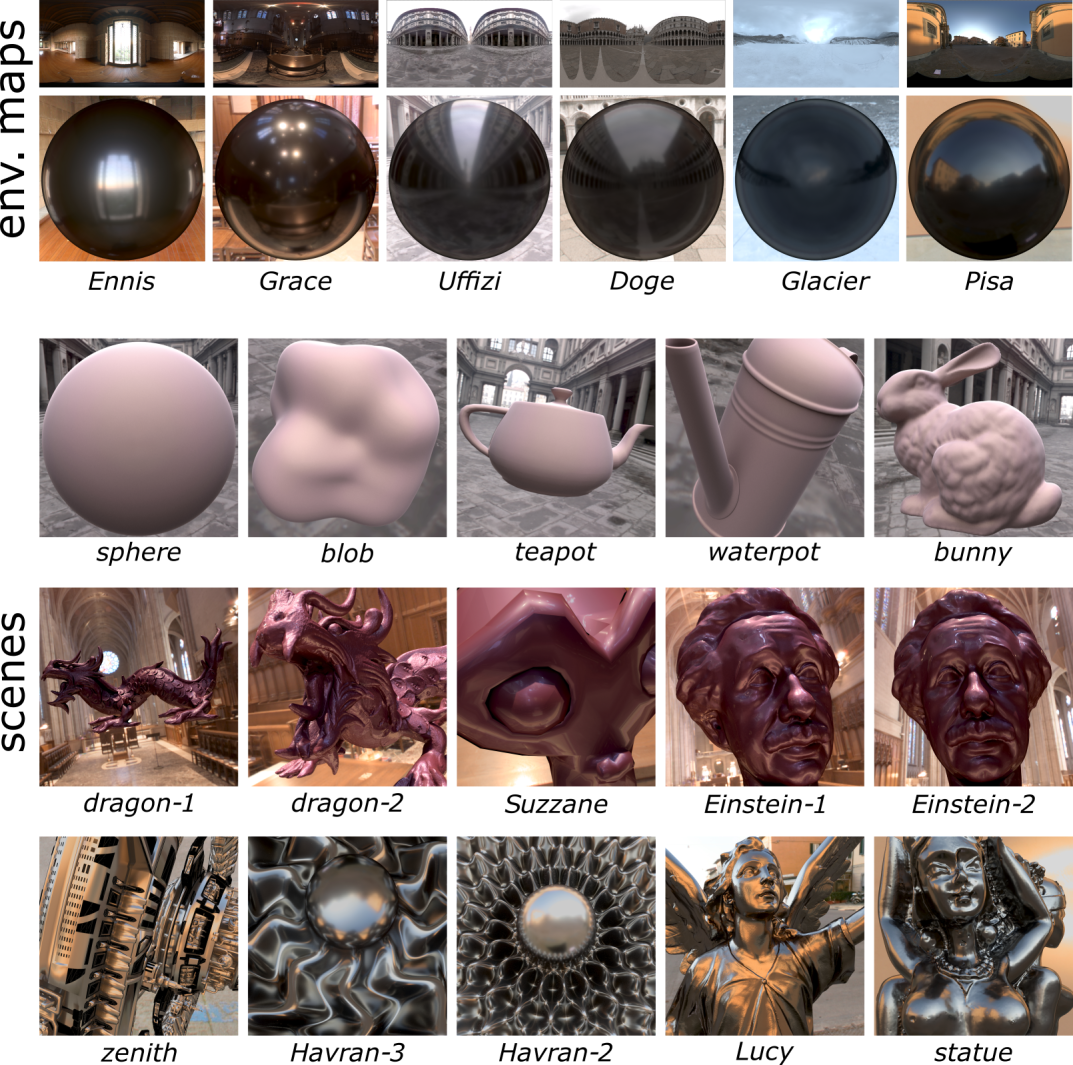} 
	\caption{\textbf{Top}: The six environment maps used in the dataset, and corresponding rendered spheres with the \emph{black-phenolic} material. \textbf{Bottom}: Sample images of all 15 scenes with different materials and illumination conditions. First row: \emph{pink-felt} and \emph{Uffizi}; second row: \emph{violet-acrylic} and \emph{Grace}; third row: \emph{nickel} and \emph{Pisa}.
	}
	
	
	\label{fig:dataset-example}
\end{figure}

By combining the aforementioned materials (100), illumination conditions (6), and scenes (15), we generate a total of 9,000 dataset samples using the Mitsuba physically-based renderer~\cite{wenzel2010}. For each one we provide: The rendered HDR image, a corresponding LDR image\footnote{ Tone-mapped using the algorithm by Mantiuk et al.~\shortcite{mantiuk2008}, with the predefined \textit{lcd office} display, and color saturation and contrast enhancement set to 1.}, along with depth, surface normals, alpha channel, and ambient occlusion maps. While not all these maps are used in the present work, we make them available with the dataset should they be useful for future research. Figure~\ref{fig:dataset-example} shows sample images for all fifteen scenes.

\section{Collecting appearance similarity information}
\label{sec:user-study}
We describe here our methodology to gather crowdsourced information about the perception of material appearance. 

\begin{figure}
\includegraphics[width=\columnwidth]{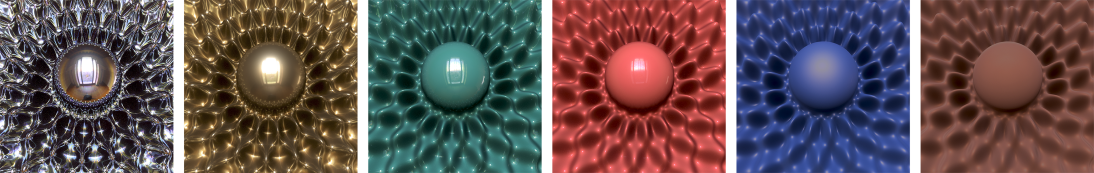}
\caption{Sample stimuli for our appearance similarity collection. They correspond to the \textit{Havran-2} scene, with materials from the MERL database, rendered with the \textit{Ennis} environment map. In reading order: \textit{chrome}, \textit{gold-metallic-paint3}, \textit{specular-green-phenolic}, \textit{maroon-plastic}, \textit{dark-blue-paint} and \textit{light-brown-fabric}.}
\label{fig:2fac_samples}
\end{figure}

\mparagraph{Stimuli} 
We use 100 different stimuli, covering all 100 materials in the dataset, rendered with the \emph{Ennis} environment map. We choose the
\emph{Havran-2} scene, since its shape has been designed to maximize the information relevant for material appearance judgements by optimizing the coverage of incoming and outgoing light directions sampled~\cite{havran2016}. Figure \ref{fig:2fac_samples} shows some examples.  

\mparagraph{Participants} A total of 603 participants took part in the test through the Mechanical Turk (MTurk) platform, with an average age of 32, and 46.27\% female.  Users were not aware of the purpose of the experiment. 

\mparagraph{Procedure} Our study deals with the \textit{perception} of material appearance, which may not be possible to represent in a linear scale; this advises against ranking methods~\cite{Kendall1940}. We thus gather data in the form of relative comparisons, following a 2AFC scheme;  the subject is presented with a triplet made up of one \emph{reference} material, and two \emph{candidate} materials, and their task is to answer the question \textit{Which of these two candidates has a more similar appearance to the reference?} by choosing one among the two candidates. This approach has several additional advantages: it is easier for humans than providing numerical distances~\cite{Mcfee2011,Schultz03}, while it reduces fatigue and avoids the need to reconcile different scales of similarity among subjects~\cite{Kendall1990}. 

However, given our 100 different stimuli, a naive 2AFC test would require 495,000 comparisons, which is intractable even if not all subjects see all comparisons. To ensure robust statistics, we aim to obtain five answers for each comparison, which would mean testing a total of 2,475,000 comparisons. Instead, we use an iterative \emph{adaptive sampling} scheme~\cite{tamuz2011adaptively}: For any given reference, each following triplet is chosen to maximize the information gain, given the preceding responses (please refer to the supplementary material for a more detailed description of the method). 
From an initial random sampling, we perform 25 iterations as recommended by Tamuz et al. for datasets our size; in each iteration we sample 10 new pairs for every one of our 100 reference materials, creating 1,000 new triplets. After this process, the mean information gain per iteration is less than $10^{-5}$, confirming the convergence of the sampling scheme. This scheme allows us to drastically reduce the number of required comparisons, while providing a good approximation to sampling the full set of triplets.

Each test (HIT in MTurk terminology) consisted of 110 triplets. To minimize worker unreliability~\cite{welinder2010multidimensional}, each HIT was preceded by a short training session that included a few trial comparisons with obvious answers~\cite{Rubinstein10Comparative,GarcesSIG2014}. In addition, ten control triplets were included in each HIT, testing repeated-trial consistency within participants. We adopt a conservative approach and reject participants with two or more different answers.
In the end, we obtained 114,840 valid answers, yielding a participants' consistency of 84.7\%.

As a separate test, to validate how well our collected answers generalize to other shapes and illuminations, we repeated the same comparisons, this time with symmetric and asymmetric triplets chosen randomly from our dataset, with the condition that they do not contain the \textit{Havran-2} shape nor the \textit{Ennis} illumination. For symmetric triplets, the three items in the triplet differ only in the material properties, while in asymmetric triplets geometry and lighting also vary. We launched 2,500 symmetric triplets, and found that participants' majority matched the previous responses with a 84.59\% rate. When we added the same number of asymmetric triplets to the test, participants' answers held with a 80\% match rate.
\section{Learning perceived similarity}
\label{sec:cnn}

This section describes our approach to learn perceived similarity for material appearance. 
Given an input image $\psi$, our model provides a feature vector $f(\psi)$ that transforms the input image into a feature space well aligned with human perception.

We use the ResNet architecture~\cite{he2015}, based on its generalization capabilities and its proven performance on image-related tasks. The novelty of this architecture is a residual block meant for learning a residual mapping between the layers, instead of a direct mapping, which enables training very deep networks (hundreds of layers) with outstanding performance.
For training we use image data from our materials dataset (Section~\ref{sec:dataset}), together with human data on perceived similarity (Section~\ref{sec:user-study}). We first describe our combined loss function, then our training procedure.

\subsection{Loss function}
\label{subsec:cnn_loss}

We train our model using a loss function consisting of two terms, equally weighted:
\begin{equation}
\mathcal L = \mathcal L_{TL} + \mathcal L_P 
\label{eq:final-perceptual-loss}
\end{equation}
The two terms represent a perceptual triplet loss, and a similarity term, respectively. The terms aim at learning appearance similarity from the participants' answers, while extracting the main features defining the material depicted in an image. In the following, we describe these terms and their contribution. 

\mparagraph{Triplet loss term $\mathcal L_{TL}$}
This term allows to introduce the collected MTurk information on appearance similarity. Let $\mathcal A = \{(r_i,a_i,b_i)\}$ be the set of answered relative comparisons, 
where $r$ is the reference image, $a$ is the candidate image chosen by the majority of users as being more similar to $r$, and $b$ the other candidate; $i$ indexes over all the relative comparisons. Intuitively, $r$ and $a$ should be closer together in the learned feature space than $r$ and $b$. 
It is not feasible to collect user answers for all possible comparisons ($n$ different images would lead to $n\binom{n-1}{2}$ tests); however, as we have shown in Section~\ref{sec:user-study}, the collected answers for a triplet $(r,a,b)$ involving materials $m^r$, $m^a$ and $m^b$ generalize well to other combinations of shape and illumination from our dataset involving the same set of materials.
%
We thus define $\mathcal A^M = \{(m^r_i,m^a_i,m^b_i)\}$ as the set of relative comparisons with collected answers ($m^a$ represents the material chosen by the majority of participants). We then formulate the first term as a triplet loss~\cite{cheng2016,schroff2015,lagunas2018}:
\begin{equation}
\mathcal L_{TL} = \frac{1}{|\mathcal B^A|}\sum_{(r, a, b) \in \mathcal B^A} \big[||f(r) - f(a) ||_2^2 - ||f(r) - f(b) ||_2^2 + \mu\big]_+ \; 
\label{eq:triplet-loss-humans}
\end{equation}
where $f(\psi)$ is the feature vector of image $\psi$, and the set $\mathcal B^A$ is defined as:
\begin{equation}
\mathcal B^A = \big[(r, a, b) \;|\; (m^{r}, m^{a}, m^{b}) \in \mathcal A^M \; \wedge \; (r, a, b) \in \mathcal B \big]
\end{equation}
%
with $\mathcal B$ the current training batch. 
In Eq.~\ref{eq:triplet-loss-humans}, $\mu$ represents the margin, which accounts for how much we aim to separate the samples in the feature space.

\mparagraph{Similarity term $\mathcal L_P$}
We introduce a second loss term that maximizes the log-likelihood of the model choosing the same material as humans. We define this probability $p_{ra}$ (and conversely $p_{rb}$) as a quotient between similarity values $s_{ra}$ and $s_{rb}$:
$$
p_{ra} = \frac{s_{ra}}{s_{rb}+ s_{ra}}\;,\quad 
p_{rb} =\frac{s_{rb}}{s_{rb}+s_{ra}}
$$
These similarities are derived from the distances between $r$, $a$ and $b$ in the feature space, where a similarity value of 1 means perfect similarity and a value of 0 accounts for total dissimilarity:
$$ 
s_{ra} = \frac{1}{1+d_{ra}}\;,\quad
s_{rb} = \frac{1}{1+d_{rb}}\;, \quad
\mathrm{where} 
$$ 
$$ 
d_{ra} = ||f(r) - f(a) ||_2^2\;,\quad
d_{rb} = ||f(r) - f(b) ||_2^2
$$ 
With this, we can formulate the similarity term as:
\begin{equation}
	\mathcal L_P = -\frac{1}{|\mathcal B^A|}\sum_{(r,a,b) \in \mathcal B^A} \log p_{ra} 
	\label{eq:perp-loss}
\end{equation}
%
%

\subsection{Training details}
\label{subsec:cnn_train}

\begin{figure}
\includegraphics[width=\columnwidth]{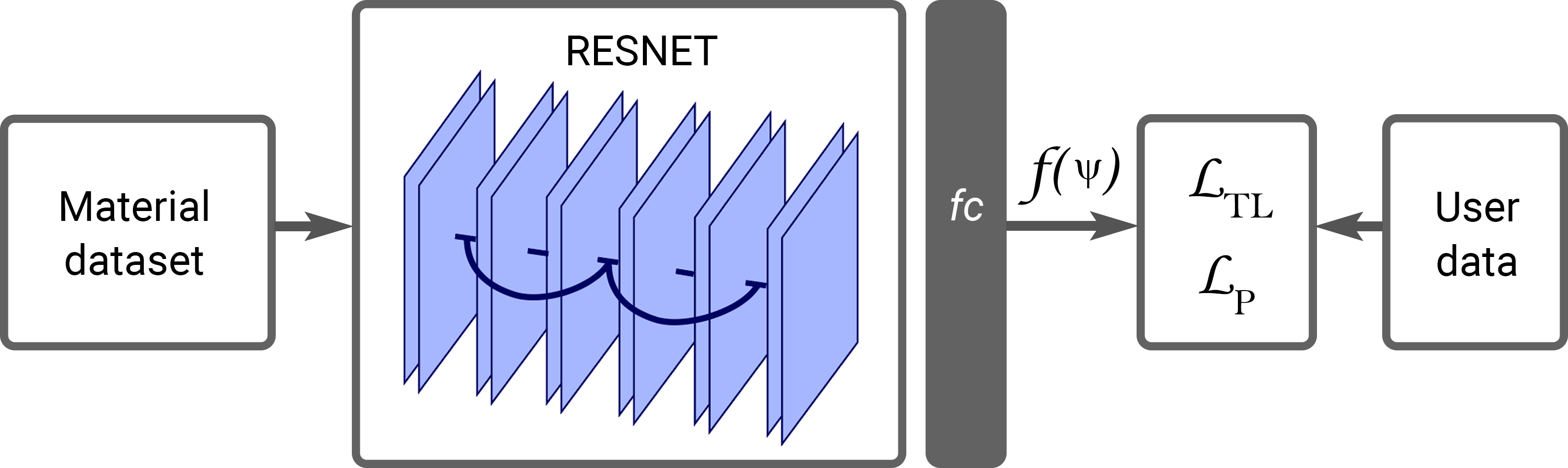}
\caption{Scheme of the training process, using both image data from our material dataset, and human data of perceived similarity. 
We train our model so that, for an input image $\psi$, it yields a 128-dimensional feature vector $f(\psi)$. 
}
\label{fig:training-scheme}
\end{figure}

For training, we remove the \textit{Havran-2} and \textit{Havran-3} scenes from the dataset, leading to 7,800 images (13 (scenes) $\times$ 6 (env. maps) $\times$ 100 (materials)), augmented to 39,000 using crops, flips, and rotations. These 39,000 images, together with the collected MTurk answers, constitute our training data. 
We use the corrected \emph{Adam} optimization~\cite{reddi2018,kingma2014} with a learning rate that starts at $10^{-3}$ to train the network. We train for 80 epochs and the learning rate is reduced by a factor of 10 every 20 epochs. For initialization, we use the weights of the pre-trained model~\cite{he2015} on ImageNet~\cite{deng2009,russakovsky2015}.
To adapt the network to our loss function, we remove the last layer of the model and introduce a fully-connected (\textit{fc}) layer that outputs a 128-dimensional feature vector $f(\psi)$. We use a margin $\mu=0.3$ for the triplet loss term $\mathcal L_{TL}$.
%
Figure~\ref{fig:training-scheme} shows a scheme of the training procedure. 

\section{Evaluation}\label{sec:results}

%

We evaluate our model on the set of images of the material dataset not used during training. We employ the \emph{accuracy} metric, which represents the percentage of triplet answers correctly predicted by our model. It can be computed as \emph{raw}, considering each of the five answers independently as the correct one, or \emph{majority}, considering the majority opinion as correct~\cite{wills2009,GarcesSIG2014}. 
Using our MTurk data from Section~\ref{sec:user-study}, the results are 73.10\% and 77.53\% respectively for human observers, indicating a significant agreement across subjects. Our model performs better than human accuracy, with 73.97\% and 80.69\% respectively. In other words, our model predicts the majority's perception of similarity almost 81\% of the time. We include an \textit{oracle} predictor in Table~\ref{tab:2afc-results1}, which has access to all the human answers and returns the majority opinion; note that its raw accuracy is not 100 due to human disagreement. Figure~\ref{fig:triplet_1} shows examples from our 26,000 queries where our model agrees with the majority response, while we discuss failure cases later in this section. More examples of queries and our model's answer are included in the supplementary material. 

\newcommand{\redthumbmain}{0.15} 
\newcommand{\seprefmain}{0.2cm} 
\newcommand{\sepmain}{0.1cm} 
\newcommand{\septripmain}{0.4cm} 

\begin{figure*}
	\centering
	\begin{tabular}{c @{\hspace{\seprefmain}}
			c @{\hspace{\sepmain}}
			c @{\hspace{\septripmain}}
			c @{\hspace{\seprefmain}}
			c @{\hspace{\sepmain}}
			c @{\hspace{\septripmain}}
			c @{\hspace{\seprefmain}}
			c @{\hspace{\sepmain}}
			c @{\hspace{\septripmain}}
			c @{\hspace{\seprefmain}}
			c @{\hspace{\sepmain}}
			c}

\includegraphics[width=\redthumbmain\columnwidth]{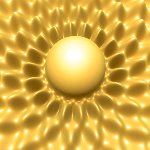} &
\includegraphics[width=\redthumbmain\columnwidth]{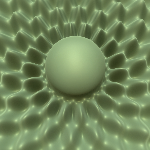}&
\includegraphics[width=\redthumbmain\columnwidth]{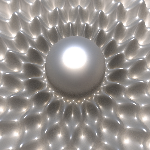}&	

\includegraphics[width=\redthumbmain\columnwidth]{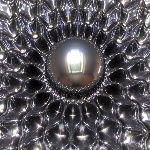} &
\includegraphics[width=\redthumbmain\columnwidth]{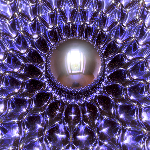}&
\includegraphics[width=\redthumbmain\columnwidth]{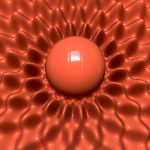}  &

\includegraphics[width=\redthumbmain\columnwidth]{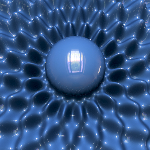} &
\includegraphics[width=\redthumbmain\columnwidth]{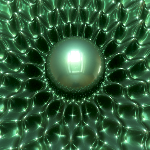}&
\includegraphics[width=\redthumbmain\columnwidth]{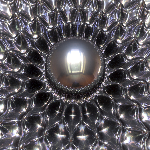}&	
	
\includegraphics[width=\redthumbmain\columnwidth]{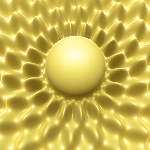} &
\includegraphics[width=\redthumbmain\columnwidth]{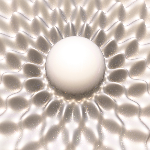}&
\includegraphics[width=\redthumbmain\columnwidth]{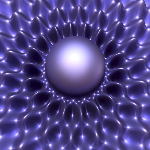}	\vspace{-0.15cm}\\

& \small \textcolor{green}{3} &	\small \textcolor{green}{2}& 
& \small \textcolor{green}{5} & \small \textcolor{green}{0}&
& \small \textcolor{green}{5} & \small \textcolor{green}{0}&
& \small \textcolor{green}{4} & \small \textcolor{green}{1} \\[1pt]

\includegraphics[width=\redthumbmain\columnwidth]{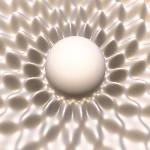} &
\includegraphics[width=\redthumbmain\columnwidth]{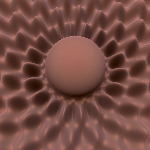}&
\includegraphics[width=\redthumbmain\columnwidth]{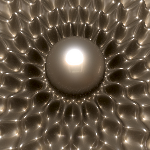}&
\includegraphics[width=\redthumbmain\columnwidth]{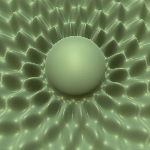} &
\includegraphics[width=\redthumbmain\columnwidth]{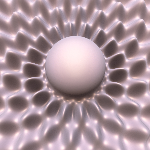}&
\includegraphics[width=\redthumbmain\columnwidth]{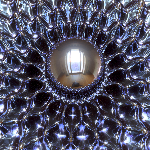} &
\includegraphics[width=\redthumbmain\columnwidth]{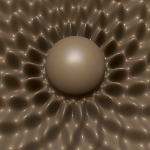} &
\includegraphics[width=\redthumbmain\columnwidth]{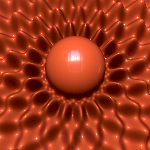}&
\includegraphics[width=\redthumbmain\columnwidth]{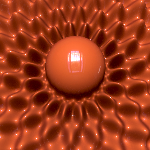}&
\includegraphics[width=\redthumbmain\columnwidth]{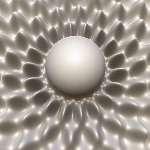} &
\includegraphics[width=\redthumbmain\columnwidth]{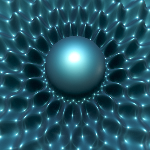}&
\includegraphics[width=\redthumbmain\columnwidth]{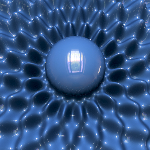} \vspace{-0.15cm}\\

& \small \textcolor{green}{5} &	\small \textcolor{green}{0}& 
& \small \textcolor{green}{5} &\small \textcolor{green}{0}&
& \small \textcolor{green}{4} & \small \textcolor{green}{1}&
& \small \textcolor{green}{3} & \small \textcolor{green}{2} \vspace{-0.15cm}\\
	
\end{tabular}	
\caption{\label{fig:triplet_1} Examples from our 26,000 queries (reference, plus the two candidates) where our model agrees with the majority response (this is the case almost 81\% of the time). The numbers indicate the number of votes each image received from the participants. More examples are included in the supplementary material. }
\end{figure*}

\mparagraph{Comparison with other metrics}
We compare the performance of our model to six different metrics used in the literature for material modeling and image similarity: The three common metrics analyzed by Fores and colleagues~\shortcite{fores2012}, the perceptually-based metrics by Sun et al.~\shortcite{sun2017} and Pereira et al.~\shortcite{pereira2012}, and SSIM \cite{wang2004image}, a well-known image similarity metric. 
We analyze again accuracy, and we additionally analyze \emph{perplexity}, which is a standard measure of how well a probability model
predicts a sample, taking into account the uncertainty in the model. Perplexity $Q$ is given by:
\begin{equation}
Q = 2 ^ {-\frac{1}{|\mathcal A|}\sum_{\Omega} \log_2 p_{ra} } 
\label{eq:perplexity}
\end{equation}
where $\Omega=(r,a)\in \mathcal A$, $|\mathcal A|$ is the number of collected answers, and $p_{ra}$ is the probability of $a$ being similar to $r$ (Section~\ref{subsec:cnn_loss}).
Perplexity gives higher weight where the model yields higher confidence; its value will be 1 for a model that gives perfect predictions, 2 for a model with total uncertainty (random), and higher than 2 for a model that gives wrong predictions. 
%
As Table~\ref{tab:2afc-results1} shows, our model captures the human perception of appearance similarity significantly better, as indicated by the higher accuracy and lower perplexity values. Note that perplexity cannot be computed for humans nor the oracle, since they are not probability distributions.

\begin{table}[t]
	\centering
	\begin{tabular}{@{}rcccccccc@{}}
		\multicolumn{5}{c}{\textsc{Evaluation of our model}} \\
		\midrule
		\multirow{2}{*}{Metric} & \multicolumn{2}{c}{Accuracy} & \multicolumn{2}{c}{Perplexity} \\ 	\cmidrule(l{2pt}r{2pt}){2-3}\cmidrule(l{2pt}r{2pt}){4-5}  
		& Raw & Majority & Raw & Majority \\ 
		\midrule
		Humans	& 73.10 & 77.53 & - & - \\	
		Oracle	& 83.79 & 100.0 & - & - \\
		\midrule		
		RMS	& 61.63 & 64.72 & 3.61 & 3.13 \\
		RMS-cos & 61.60 & 64.67 & 3.86 & 3.33 \\
		Cube-root & 63.71 & 67.40 & 1.96 & 1.86 \\
		L2-lab	& 63.76 & 67.21 & 2.16 & 2.07 \\
		L4-lab	& 60.60 & 62.93 & 15.36 & 11.66 \\
		SSIM & 62.35 & 64.74 & 2.02 & 1.94 \\
		\midrule 
		\textbf{Our model} & 73.97 & 80.69 & 1.74 & 1.55 \\
		\midrule 
	\end{tabular}
	\caption{Accuracy and perplexity of our model compared to human performance, an oracle (which always returns the majority opinion), and six other metrics from the literature: RMS, RMS-cos, Cube-root~\cite{fores2012}, L2-lab~\cite{sun2017}, L4-lab~\cite{pereira2012} and SSIM \cite{wang2004image}.  For accuracy, higher values are better, while for perplexity lower are better. }
	\label{tab:2afc-results1}
\end{table}

\begin{figure}
	\includegraphics[width=\columnwidth]{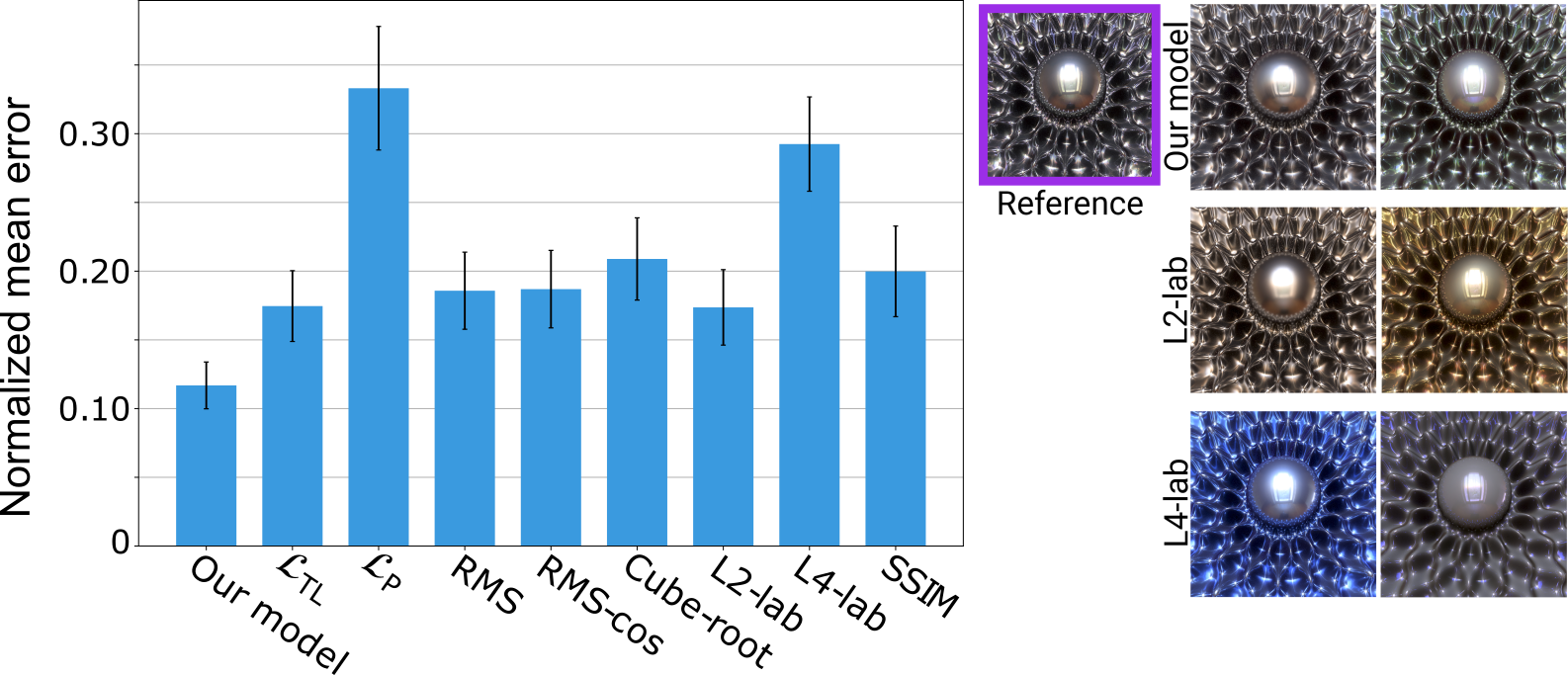}
	\caption{\textbf{Left}: Mean error for different metrics (each normalized by its maximum value) with respect to distances derived from human responses,	
	across all possible pair combinations from the MERL dataset (the $\mathcal L_{TL}$ and $\mathcal L_{P}$ columns refer to the ablation studies in Table~\ref{tab:ablation-studies}; please refer to the main text). Error bars correspond to a $95\%$ confidence interval. \textbf{Right:} Representative example of the two most similar materials to a given reference, according to (from top to bottom): Our model, and the two perceptually-based metrics L2-lab~\cite{sun2017}, and L4-lab~\cite{pereira2012}. Our model yields less error, and captures the notion of appearance similarity better.  }
	\label{fig:other_distances}
\end{figure}

Additionally, we compute the mean error between
distances derived from human responses and our model's predictions, across all possible material pair combinations from the MERL dataset. To obtain the derived distances from the collected human responses, we use t-Distributed Stochastic Triplet Embedding (tSTE)~\cite{van2012stochastic}, which builds an n-dimensional embedding that aims to correctly represent participants' answers. We use a value of $\alpha=5$ (degrees of freedom of the Student-t kernel), which correctly models $87.36\%$ of the participants' answers. We additionally compute the mean error for
the six other metrics. As shown in Figure~\ref{fig:other_distances}, our metric yields the smallest error. 
Error bars correspond to a $95\%$ confidence interval.

\mparagraph{Ablation study}
We evaluate the contribution of each term in our loss function to the overall performance via a series of ablation experiments (see Table~\ref{tab:ablation-studies}). We first evaluate performance using only one of the two terms ($\mathcal L_{TL}$ and $\mathcal L_{P}$) in isolation.
We also analyze the result of incorporating two additional loss terms, which could in principle apply to our problem: A cross-entropy term $\mathcal L_{CE}$, and a batch-mining triplet loss term $\mathcal L_{BTL}$. The former aims at learning a soft classification task by penalizing samples which do not belong to the same class~\cite{szegedy2015}, while the latter has been proposed in combination with the cross-entropy term to improve the model's generalization capabilities and accuracy~\cite{gao2018revisiting} (more details about these two terms can be found in the appendix). Last, we analyze performance using \textit{only} these two terms ($\mathcal L_{CE}$ and $\mathcal L_{BTL}$), without incorporating participants' perceptual data.
As Table~\ref{tab:ablation-studies} shows, none of these alternatives outperforms our proposed loss function. Although the single-term $\mathcal L_{P}$ loss function yields higher accuracy, it also outputs higher perplexity values; moreover, as Figure~\ref{fig:other_distances} shows, the mean error is much higher, meaning that it does not capture the notion of similarity as well as our model.


\small{
\begin{table}[t]
\centering
\begin{tabular}{@{}rcccccccc@{}}
\multicolumn{5}{c}{\textsc{Ablation study and alternative networks}} \\
\midrule
\multirow{2}{*}{Model} & \multicolumn{2}{c}{Accuracy} & \multicolumn{2}{c}{Perplexity} \\ \cmidrule(l{2pt}r{2pt}){2-3}\cmidrule(l{2pt}r{2pt}){4-5}  
& Raw & Majority & Raw & Majority \\ 
\midrule
$\mathcal L_{TL}$ & 69.32 & 74.12 & 1.89 & 1.73 \\
$\mathcal L_{P}$ & 75.22 & 82.31 & 3.16 & 2.13 \\
$\mathcal L_{TL} + \mathcal L_{P}+ \mathcal L_{CE}$ & 71.82 &  77.53 & 1.76 & 1.66 \\
$\mathcal L_{TL} + \mathcal L_{P}+ \mathcal L_{CE}+ \mathcal L_{BTL}$ & 71.78 & 77.76 & 1.76 & 1.67 \\
$\mathcal L_{CE} + \mathcal L_{BTL}$ & 56.88 & 58.44 & 1.96 & 1.93 \\
\midrule
VGG & 70.70 & 76.40 & 2.25 & 1.89 \\
DenseNet & 60.90 & 63.49 & 2.66 & 2.46 \\
\midrule
\textbf{Our model} & 73.97 & 80.69 & 1.74 & 1.55 \\
\midrule	
\end{tabular}
\caption{Accuracy and perplexity for other loss functions, as well as for two alternative architectures (VGG and DenseNet).}
\label{tab:ablation-studies}
\end{table}
}

\normalsize 

\mparagraph{Alternative networks}
We have tested two alternative architectures, VGG~\cite{simonyan2014}, which stacks convolutions with non-linearities; and DenseNet~\cite{huang2017}, which introduces concatenations between different layers. Both models have been trained using our loss function. As shown in Table~\ref{tab:ablation-studies}, both yield inferior results compared to our model. DenseNet has a low number of learned parameters, insufficient to capture the data distribution, hampering convergence. VGG has a larger number of parameters; however, the residual mapping learned by the residual blocks in the architecture of our model yields the best overall performance.


\mparagraph{Results by category}
We additionally divide the materials into eight categories: \textit{acrylics, fabrics, metals, organics, paints, phenolics, plastics}, and \textit{other}, and  analyze raw and majority accuracy in each. We can see in Table~\ref{tab:2afc-results} how our model is reasonably able to predict human perception also within each category. For instance, although the numbers are relatively consistent across all the categories, humans perform on average slightly worse for phenolics or acrylics, and better for fabrics; our metric mimics such behavior. The only significant difference occurs within the \textit{organics} category, where our metric performs worse than humans. 
This may be due to the combination of a low number of material samples and a large variety of appearances within such category, which may hamper the learning process.

\begin{table*}[h!]
	\centering
	\begin{tabular}{@{}r|c|c|cc|cc|cc|c@{}}
		\multicolumn{9}{c}{\textsc{Analysis per material category}} \\
		\midrule
		\multirow{2}{*}{Category} & \multirow{2}{*}{Materials} & \multirow{2}{*}{Answers} & \multicolumn{2}{c|}{Humans} & \multicolumn{2}{c|}{\textbf{Our model}} & \multicolumn{2}{c}{Oracle} \\ 	
		\cmidrule(l{5pt}r{5pt}){4-5} \cmidrule(l{5pt}r{5pt}){6-7} \cmidrule(l{5pt}r{5pt}){8-9} 
		& & & Raw & Majority & Raw & Majority & Raw \\ 
		\midrule
		Acrylics	&  4 & 4719 & 67.27 & 70.69 & 67.57 & 74.18 & 79.89\\
		Fabrics		& 14 & 16019 & 79.65 & 83.70 & 83.03 & 90.44 & 87.87 \\
		Metals		& 26 & 32337 & 74.20 & 78.90 & 75.63 & 83.10 & 84.54 \\
		Organics	&  7 & 8370 & 69.28 & 73.08 & 60.46 & 62.43 & 81.28 \\
		Paints		& 14 & 15101 & 74.22 & 78.85 & 75.22 & 81.84 & 84.61 \\
		Phenolics	& 12 & 13025 & 66.49 & 70.53 & 67.62 & 74.36 & 79.72 \\
		Plastics	& 11 & 12031 & 70.53 & 74.70 & 69.25 & 74.06 & 82.05 \\
		Other		& 12 & 13198 & 74.80 & 79.38 & 78.21 & 86.11 & 84.89 \\
		\midrule 
		Total		& 100 & 114800 & 73.10 & 77.53 & 73.97 & 80.69 & 83.79 \\
		\midrule 
	\end{tabular}
	\caption{Statistics per category. \textbf{From left to right:} Category, number of materials in each category, number of collected answers, humans' accuracy (raw and majority), accuracy of our model, and oracle raw accuracy. }
	\label{tab:2afc-results}
\end{table*}

\begin{figure}
     \includegraphics[width=\columnwidth]{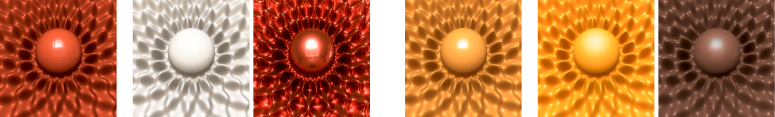}
     \caption{Two examples where humans' majority disagrees with our metric. For both, humans agreed that the middle stimulus is perceptually closer to the reference on the left, while our metric scores the right stimuli as more similar. }
     \label{fig:failure}
\end{figure}

\mparagraph{Failure cases}
Being on par with human accuracy means that our similarity measure disagrees with the MTurk majority 19.31\% of
the time. Figure~\ref{fig:failure} shows two examples where humans were consistent in choosing one stimuli as closer to the reference (5 votes out of 5), yet our metric predicts that the second one is more similar. In the leftmost example, the softness of shadows may have been a deciding factor for humans. In the rightmost example, 
humans may have been overly influenced by color, whilst our metric has factored in the presence of strong highlights. These examples are interesting since they illustrate that neither color nor reflectance are persistently the dominant factors when humans judge appearance similarity between materials.

%

\section{Applications}
\label{sec:apps}

We illustrate here several applications directly enabled by our similarity measure.

\begin{figure*}[ht]
	\includegraphics[width=0.95\textwidth]{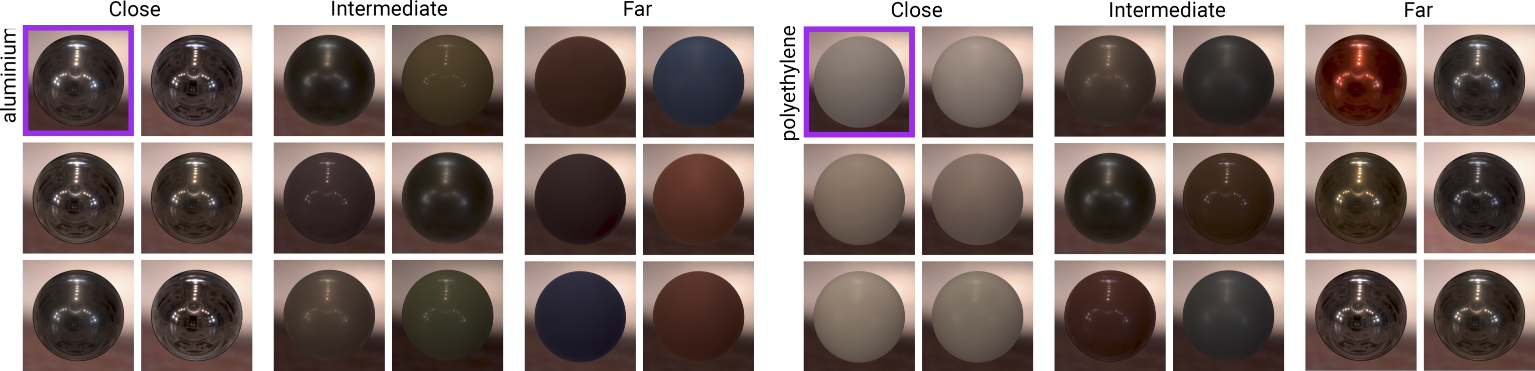}
	\caption{Two examples of material suggestions using our model. Queries from MERL (violet frame), and returned results for perceptually close, intermediate, and far away materials from the Extended MERL dataset.}
	\label{fig:apps-suggestion-01}
\end{figure*}

\mparagraph{Material suggestions} Assigning materials to a complex scene is a laborious process~\cite{Zsolnai2018,chen2015magic}. We can leverage the fact that the distances in our learned feature space correlate with human perception of similarity to provide controllable material suggestions. The artist provides the system with a reference material, and the system delivers perceptually similar (or farther away) materials in the available dataset, thus creating a controlled amount of variety without the burden of manually selecting each material. 
Figure~\ref{fig:teaser} illustrates this, where the search distance is progressively extended from a chosen reference, and the materials are then assigned randomly to each cube. 
Suggestions need not be automatically assigned to the models in the scene, but may also serve as a palette for the artist to choose from, facilitating browsing and navigation through material databases. 
Figure~\ref{fig:apps-suggestion-01} shows two MERL samples used as queries, along with returned suggestions from the \emph{Extended MERL} dataset~\cite{serrano2016}. The figure shows results at close, intermediate, and far distances from the query. 
Additional examples can be seen in Figure~\ref{fig:apps-suggestion-02}, and in the supplementary material. 

%
\begin{figure*}[ht]
	\includegraphics[width=0.95\textwidth]{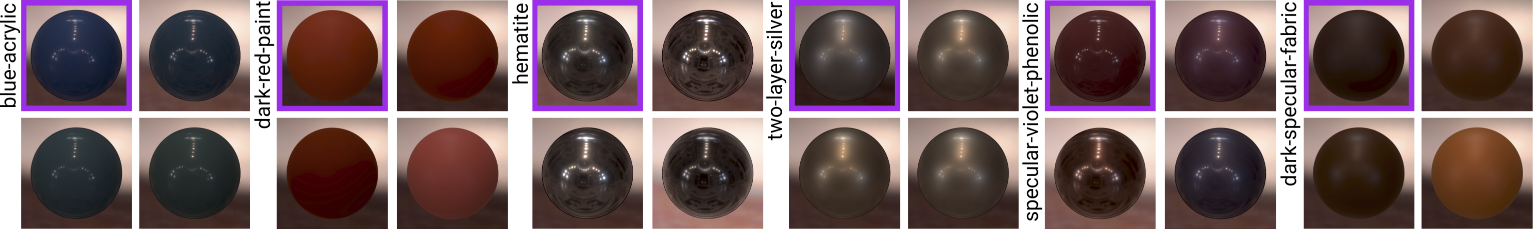}
	\caption{Additional material suggestion results. Queries (violet frame) and results for the closest materials in the \emph{Extended MERL} dataset.}
	\label{fig:apps-suggestion-02}
\end{figure*}

\begin{figure}[t]
	\includegraphics[width=\columnwidth]{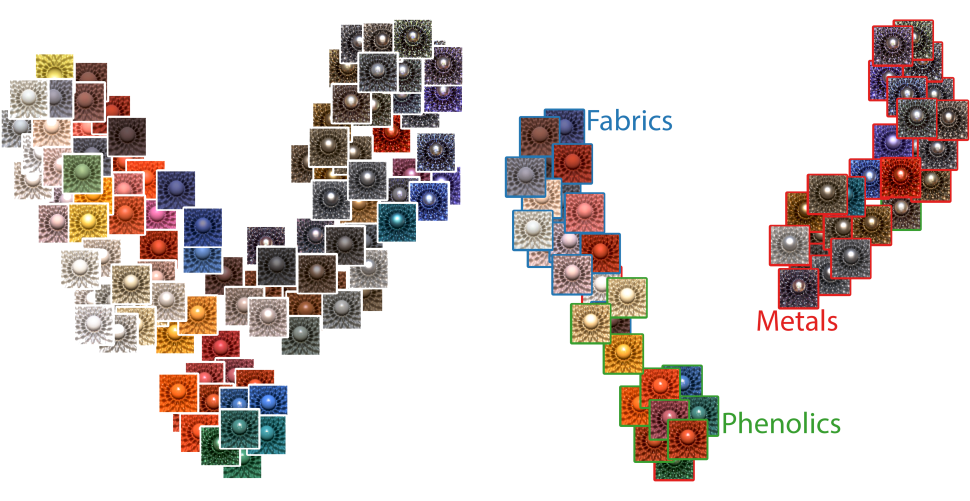} 
	\caption{Visualization of the MERL dataset in a 2D space based on the feature vectors provided by our model, using UMAP~\cite{mcinnes2018umap}. \textbf{Left:} The entire MERL dataset. \textbf{Right:} Materials from three different categories (\emph{metals}, \emph{fabrics}, and \emph{phenolics}).}
	\label{fig:apps-umap-merl}
\end{figure}

\mparagraph{Visualizing material datasets} The feature space computed by our model can be used to visualize material datasets in a meaningful way, using dimensionality reduction techniques. We illustrate this using UMAP (Uniform Manifold Approximation and Projection~\cite{mcinnes2018umap}), which helps visualization by preserving the global structure of the data. Figure~\ref{fig:apps-umap-merl} shows two results for the MERL dataset, using images not included in the training set. On the left, we can observe a clear gradient in reflectance, increasing from left to right, with color as a secondary, softer grouping factor. The right image shows a similar visualization using only three categories:  \emph{metals}, \emph{fabrics}, and \emph{phenolics}.

\begin{figure*}[ht]
	\includegraphics[width=\textwidth]{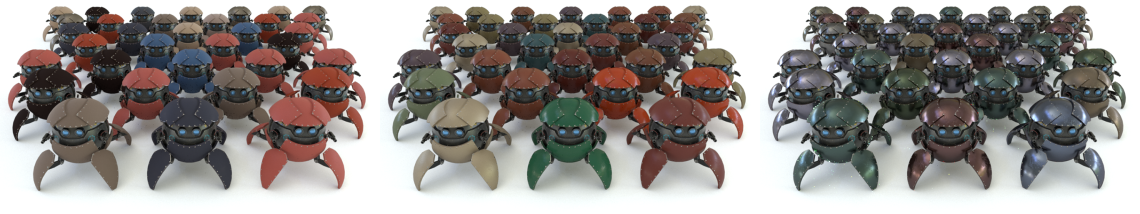}
	\caption{Material suggestions using our perceptual database clustering. The images show random materials assigned from three different clusters of varying appearance.}
	\label{fig:robots}
\end{figure*}

\begin{figure}[ht]
	\includegraphics[width=\columnwidth]{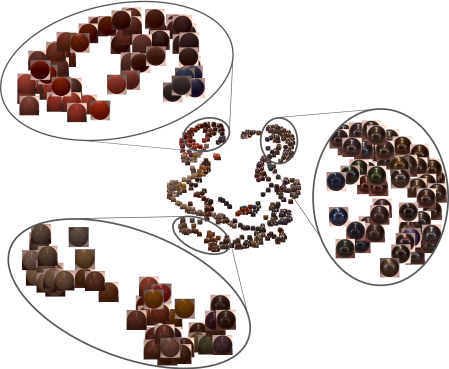}
	\caption{Visualization of the Extended MERL dataset in a 2D space based on the feature vectors provided by our model, using UMAP. Close-ups illustrate how materials of similar appearance are clearly clustered together. A larger version is included in the supplementary material.}
	\label{fig:apps-umap-extmerl}
\end{figure}

\mparagraph{Database clustering}
For unlabeled datasets like Extended MERL, our feature space allows to obtain clusters of perceptually similar materials, as shown in Figure~\ref{fig:apps-umap-extmerl} using UMAP. The close-ups highlight how materials with similar appearance are correctly grouped together by our model. 
To further analyze the clustering enabled by our perceptual feature space, we rely on the Hopkins statistic, which estimates randomness in a data set~\cite{banerjee2004validating}. A value of 0.5 indicates a completely random distribution, lower values suggest regularly-spaced data, and higher values (up to a maximum of 1) reveal the presence of clusters. 
The Hopkins statistic computed over our 128-dimensional feature vectors for the Extended MERL dataset yields a value of 0.9585, suggesting that meaningful clusters exist in our learned feature space\footnote{ This is an averaged value over 100 iterations, since the computation of the Hopkins statistic involves random sampling of the elements in the dataset.}. For comparison purposes, using only \textit{metals} in MERL the Hopkins statistic drops to 0.6935, since their visual features are less varied within that category. Figure~\ref{fig:robots} shows an example of material suggestions leveraging our perceptual clusters in unlabeled datatsets. 
\begin{figure}[ht]
	\includegraphics[width=0.95\columnwidth]{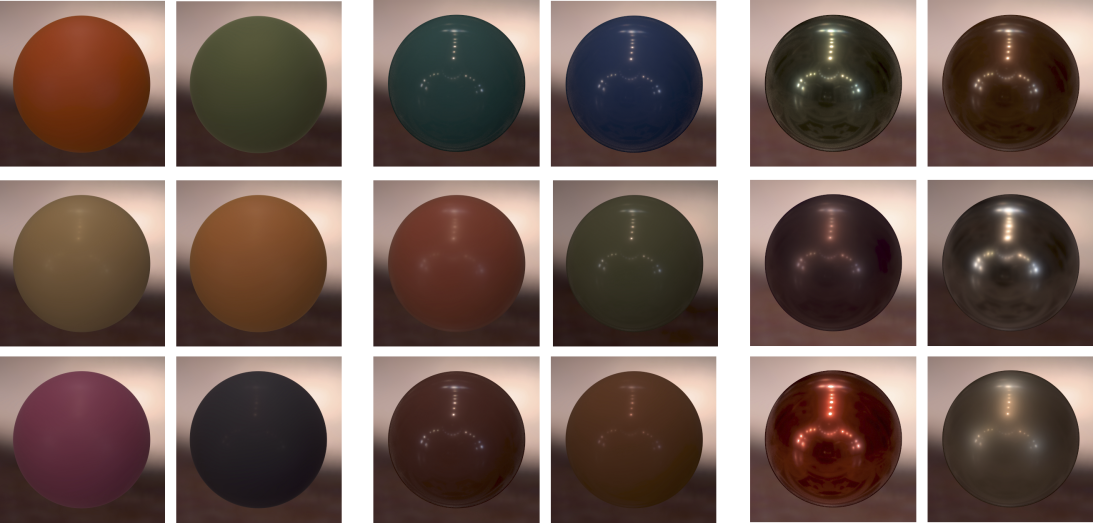}
	\caption{Representative samples of three clusters on the Extended MERL database. The Hopkins statistic on our feature space confirms that our similarity metric creates perceptually-meaningful clusters of materials.}
	\label{fig:apps-clustering}
\end{figure}

\begin{figure}[ht]
	\includegraphics[width=0.95\columnwidth]{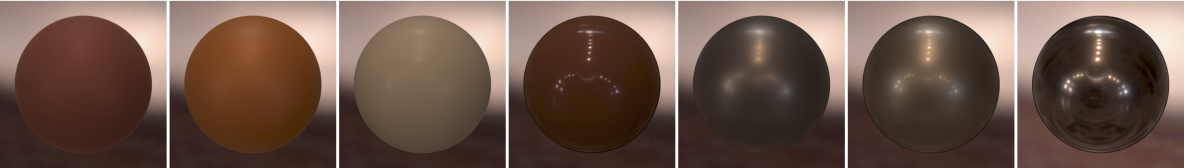}
	\caption{Example of database summarization for the Extended MERL dataset. These seven samples represent the variety of material appearances in the dataset.}
	\label{fig:apps-summarization}
\end{figure}

\mparagraph{Database summarization}
Perceptually meaningful clustering leads in turn to the possibility of database summarization. We can estimate the appropriate number of clusters using the elbow method, taking the number of clusters that explains the $95\%$ of the variance in our feature vectors. In the 400-sample Extended MERL dataset, this results in seven clusters. Taking the closest material to the centroid for each one leads to a seven-sample database summarization that represents the variety of material appearances in the dataset (Figure~\ref{fig:apps-summarization}).

%

\begin{figure}[th]
	\includegraphics[width=0.95\columnwidth]{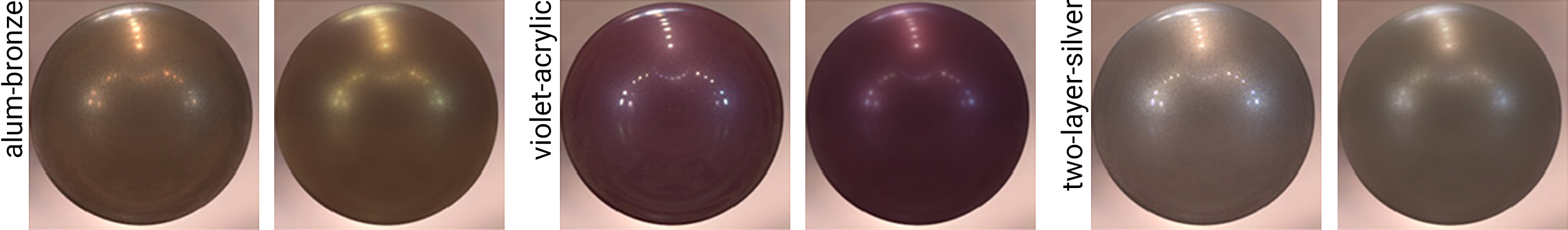}
	\caption{Our similarity metric can be used for gamut mapping applications, by minimizing the perceptual distance of our feature vectors. Each pair shows the ground truth (left), and our in-gamut result (right).}
	\label{fig:gamut}
\end{figure}

\mparagraph{Gamut mapping}
In general, our model can be used for tasks that involve minimizing a distance. This is the case for instance of gamut mapping, where the goal is to bring an out-of-gamut material into the available gamut of a different medium, while preserving its visual appearance; this is a common problem with current printing technology, or in the emerging field of computational materials. We illustrate the effectiveness of our technique in the former. Gamut mapping can be formulated as a minimization on image space \cite{pereira2012,sun2017}. We can use our feature vector $f(\psi)$ to minimize the perceptual distance between two images as $min_w ||f(o) - f(g*w)||_2^2$, where $o$ is the out-of-gamut image, and $g*w$ represents the image in the printer's gamut, defined as a linear combination of inks $g$~\cite{Matusik2009PSR}). Figure~\ref{fig:gamut} shows some examples.

\section{Discussion}
\label{sec:discussion}

\begin{figure}[ht]
	\includegraphics[width=0.9\columnwidth]{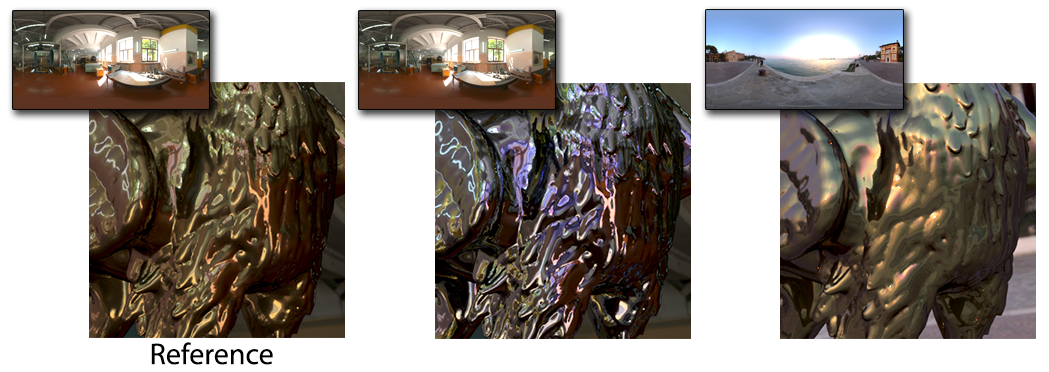}
	\caption{
	In the feature space defined by our model, the middle image (\emph{chrome}) is closer in appearance to the reference (\emph{brass}) than the image on the right (\emph{brass}). The insets show the environment maps used. Our model is driven by appearance similarity, and does not attempt to classify materials.}
	\label{fig:dist_ref_color_geom_ilum}
\end{figure}

We have presented and validated a model of material appearance similarity that correlates with the human perception of similarity. Our results suggest that a shared perception of material appearance does exist, and we have shown a number of applications using our metric.
Nevertheless, material perception poses many challenges; as such there are many exciting topics not fully investigated in this work. 
Several factors come into play that influence material appearance, i.e., the visual impression of a material, in a highly complex manner; fully identifying them and understanding their complex interactions is an open, fundamental problem. 
%
%
%
As a consequence of these interactions, the same material (e.g., plastic) may have very diverse visual appearances, whereas two samples of the same material may look very different under different illumination conditions~\cite{vangorp2007,fleming2003real}. In aiming for material appearance similarity, we aim for a material similarity metric that can predict human judgements. There is a distinction, common in fields like psychology or vision science, between the distal stimulus---the physical properties of the material---, and the proximal stimulus---the image that is the input to perception---. The key observation here is that human perceptual judgements usually lie between these two, and our training framework and loss function are designed to take both into account. We combine the information about the physical properties of the material contained in the images, by having the same material under different geometries and illuminations, with the human answers on appearance similarity. In other words, a pure image similarity metric would not be able to generalize across shape, lighting or color, while a BRDF-based metric would be unable to predict human similarity judgements.
We do not attempt to identify nor classify materials (Figure~\ref{fig:dist_ref_color_geom_ilum}). 
Our loss function could, however, incorporate additional terms (such as the cross-entropy and batch-mining triplet loss term discussed in the appendix) to help with classification tasks. We have carried out some tests and found anecdotical evidence of this, but a thorough analysis requires a separate study not covered in this work.%

\begin{figure}[ht]
	\includegraphics[width=\columnwidth]{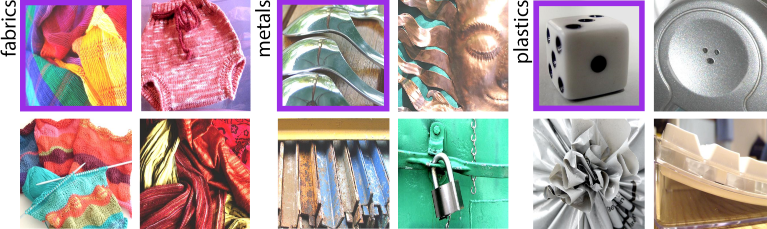}
	\caption{Results using highly heterogeneous materials from the FMD dataset. We show the three closest results returned by our model, from the reference materials highlighted in violet. Note that the search was performed across all three categories shown, not within each category. }
	\label{fig:other_datasets}
\end{figure}

Despite having trained our model on isotropic materials, we have found that it may also yield reasonable results with higher-dimensional inputs. 
Figure~\ref{fig:other_datasets} shows three examples from the Flickr Material Database (FMD)~\cite{sharan2009}, which contains captured images of highly heterogeneous materials. We have gathered all the materials from the \textit{fabrics, metals}, and \textit{plastics} categories in the database; taking one reference from each, we show the three closest results returned by our model, using an L2 norm distance in feature space. Images were resized to match the model's input size, with no further preprocessing. Note that the search was not performed within each category but across all three, yet our model successfully finds similar materials for each reference. This is a remarkable, promising result; however, a more comprehensive analysis of in-the-wild, heterogeneous materials is out of the scope of this paper.

We have also tested the performance of our model on grayscale images. In this case, we have repeated the evaluation conducted in Table~\ref{tab:2afc-results1} for our model, using grayscale counterparts of the images. Despite the removal of color information, we obtain results similar to those of our model on color images: A raw accuracy of 72.55 (vs 73.97 on color images), a majority accuracy of 78.64 (vs 80.69), a raw perplexity of 1.82 (vs 1.74), and a majority perplexity of 1.67 (vs 1.55). This further enforces the idea that we learn a measure of appearance similarity, and not image similarity.
%

To collect similarity data for material appearance, we have followed an adaptive sampling scheme~\cite{tamuz2011adaptively}; following a different sampling strategy may translate into additional discriminative power and further improve our results.
Our model could potentially be used as a feature extractor, or as a baseline for transfer-learning~\cite{sharif2014cnn, yosinski2014transferable} in other material perception tasks. A larger database could translate into an improvement of our model's predictions; upcoming databases of complex measured materials (e.g., Dupuy et al.~\shortcite{dupuy2018}) could be used to expand our training data and lead to a richer and more accurate analysis of appearance. Our methodology for data collection and model training could be useful in these cases. Similarly, upcoming network architectures that may outperform our ResNet choice could be adopted within our framework. Finding hand-engineered features could also be an option and may increase interpretability, but it could also introduce bias in the estimation.

In addition to the applications we have shown, we hope that our work can inspire additional research and different applications. For instance, our model could be of use for designing computational fabrication techniques that take into account perceived appearance. It could also be used as a distance metric for fitting measured BRDFs to analytical models, or even to derive new parametric models that better convey the appearance of real world materials. We have made our data available for further experimentation, in order to facilitate the exploration of all these possibilities.

\appendix

\section{Additional loss terms}

We describe here the two additional loss terms that we evaluate in our ablation study (refer to Section~\ref{sec:results} for details).

\mparagraph{Cross-entropy term $\mathcal L_{CE}$}
This term accounts for the soft-label cross entropy~\cite{szegedy2015}. It aims at learning a soft classification task by penalizing samples which do not belong to the same class. In our case, each material represented in the dataset can constitute a class, and the set of classes in the dataset is $\mathcal K$. Given an image $r$ included in a training batch $\mathcal B$, the probability of $r$ belonging to a certain class $k \in \mathcal K$ is given by $p_k(r)$. The cross-entropy loss term is given by:
\begin{gather}
\mathcal L_{CE} = \frac{1}{|\mathcal B|} \sum_{r \in \mathcal B} s(r)\\
s(r) = - \sum_{k\in \mathcal K} \big[ (1 - \epsilon)\log p_k(r)l_k(r) + \epsilon\log p_k(r)u(k) \big]
\label{eq:cross-entropy}
\end{gather}
where $l(r)$ is the one-hot encoding of the ground truth label, and $l_k(r)$ is the value of the vector for label $k$ (note that our training image data can be labeled, since it comes from the materials dataset presented in Section~\ref{sec:dataset}). The value of $\epsilon$ is set to 0.1, and we use the uniform distribution $u(k) = \frac{1}{|\mathcal K|}$. Both $\epsilon$ and $u(k)$ work as regularization parameters so that a wrong prediction does not penalize the cost function aggressively, while preventing overfitting.

\mparagraph{Batch-mining triplet loss term ($\mathcal L_{BTL}$)}
%
In learned models for classification or recognition, a batch-mining triplet loss has been proposed in combination with a soft-label cross entropy term such as the one we use to improve the model's generalization capabilities and accuracy~\cite{gao2018revisiting}. It is modeled as:
\begin{multline}
\mathcal L_{BTL} =  \frac{1}{|\mathcal B|} \sum_{r \in \mathcal B}\big[\underset{x^+_i}{\operatorname{argmax}} \big(||f(r) - f(x^+_i)||_2^2\big) \\
- \underset{x^-_i}{\operatorname{argmin}} \big(||f(r) - f(x^-_i)||_2^2 \big) + \mu \big]_+
\label{eq:triplet-loss}
\end{multline}
where $x^+_i$ designates images of the training batch $\mathcal B$ belonging to the same class as $r$, and $x^-_i$ images belonging to a different class than $r$. Intuitively, this loss mines and takes into consideration the hardest examples within each batch, improving the learning process.



\begin{acks}
	We want to thank the anonymous reviewers for their encouraging and insightful feedback on the manuscript, Ibon Guillen for his invaluable help using Mitsuba; Miguel Galindo, Ibon Guillen, Adrian Jarabo, and Julio Marco for their help setting up the scenes, and the members of the Graphics and Imaging Lab for the discussions about the paper. Sandra Malpica was supported by a DGA predoctoral grant (period 2018-2022). Ana Serrano was supported by an FPI grant from the Spanish Ministry of Economy and Competitiveness, and a Nvidia Graduate Fellowship. Elena Garces is additionally supported by a Juan de la Cierva Fellowship. This project has received funding from the European Research Council (ERC) under the European Union's Horizon 2020 research and innovation programme (CHAMELEON project, grant agreement No 682080) and the Spanish Ministry of Economy and Competitiveness (projects TIN2016-78753-P, and TIN2016-79710-P).
\end{acks}

\bibliographystyle{ACM-Reference-Format}
\bibliography{bibliography}

\end{document}